\documentclass[conference]{IEEEtran}
\IEEEoverridecommandlockouts
\usepackage{cite}
\usepackage{amsmath,amssymb,amsfonts}
\usepackage{algorithmic}
\usepackage{graphicx}
\usepackage{textcomp}
\usepackage{xcolor}
\usepackage{hyperref}
\hypersetup{hidelinks, breaklinks}
\usepackage{doi}

\def\BibTeX{{\rm B\kern-.05em{\sc i\kern-.025em b}\kern-.08em
    T\kern-.1667em\lower.7ex\hbox{E}\kern-.125emX}}
\usepackage{breakurl}

\begin{document}

\title{Barriers to implementation of blockchain technology in agricultural supply chain}

\author{\IEEEauthorblockN{David Cuellar
}
\IEEEauthorblockA{\textit{Dept. Computing \& Informatics} \\
\textit{Bournemouth University}\\
\textit{Bournemouth, United Kingdom}\\
\textit{s5535743@bournemouth.ac.uk}}
\and
\IEEEauthorblockN{Zechariah Johnson 
}
\IEEEauthorblockA{\textit{Dept. Computing \& Informatics} \\
\textit{Bournemouth University}\\
\textit{Bournemouth, United Kingdom}\\
\textit{s5526256@bournemouth.ac.uk}}
}

\maketitle

\begin{abstract}
Emerging technologies, such as Blockchain and the Internet of Things (IoT), have had an immense role in propelling the agricultural industry towards the fourth agricultural revolution. Blockchain and IoT can greatly improve the traceability, efficiency, and safety of food along the supply chain. Given these contributions, there are many barriers to widespread adoption of this technology, including a deficit in many workers’ ability to understand and effectively use this technology in addition to a lack of infrastructure to educate and support these workers. This paper discusses the barriers to adoption of blockchain and IoT technology in the agricultural supply chain. The authors analyse the impact of Blockchain and IoT in the food supply chain and methods in which governments and corporations can become more adaptable. Through the reduction in imports and protection of demand for local farmers, developing economies can create local sustainable agricultural ecosystems. Furthermore, the use of both public and private Research and Development can greatly contribute to the global knowledge on new technologies and improve many aspects of the food supply chain. In conclusion, both governments and corporations have a big role to play in the increased implementation of progressive technologies and the overall improvement of the food supply chain along with it. 
\end{abstract}


\section{Introduction}

Blockchain technology has been responsible for recent innovations in many industries. Blockchain can be defined as a chain of blocks that store transactions in a decentralized ledger \cite{ASurveyOfBlockchain}. The use of blockchain would be for any network seeking “decentralization, immutability, transparency, and auditability” \cite[p. 117134]{ASurveyOfBlockchain}. The internet of things (IoT) is characterized as a network of things, connected wirelessly through smart sensors \cite{TheInternetOfThings}. Developments in these technologies have led to many breakthroughs in the agriculture industry and supply chain, although many barriers still exist that prevent widespread adoption. This paper aims to explore the barriers preventing the use of blockchain technology, along with the use of IoT, in the agricultural supply chain. 

With food safety and food security being among the most important problems in today's world, the literature review below will demonstrate that there is a great need for efficiency and transparency along the food supply chain. The literature review will also describe the gap between the empowering technologies available and the skills required in workers to take full advantage of them. This paper will first discuss blockchain and IoT in improving the process of traceability along the food supply chain, leading to faster and safer food for those that need it. Furthermore, this paper advocates for reform in public policy and Research \& Development (R\&D) to create lasting systems of technological implementation across the food supply chain.  

The rest of this paper will be formatted as follows: first, a literature review of relevant work. This review will begin with the exploration of traceability alongside other aspects of blockchain and IoT used in agriculture. The literature review will also analyse industry 4.0 and the ways in which workers are adapting to it. Next, the authors will provide a research rationale to explain the issue at hand and why it is worth studying. Following this will be the author's proposed solution: the use of public policy and R\&D to increase the adoption of blockchain and IoT in the food supply chain. The authors will then provide an analysis of the strengths and weaknesses of this proposed solution. The paper will conclude with a call for future research and a discussion of the findings.

\section{Literature review}

\subsection{Blockchain technology in supply chain}

    Although the main application is traceability, it is necessary to provide a comprehensive view on the applications of blockchain technology in agriculture. There are numerous applications, however, Yadav and
    Singh \cite{ASystematicLiterature} suggest that the main ones are:

    \subsubsection{Traceability} Olsen and Borit \cite[p. 148]{HowToDefineTraceability} define traceability as “The ability to access any or all information relating to that which is under consideration, throughout its entire life cycle, by means of recorded identifications”. It can be concluded that, in the field of agriculture, the more accurate the data on the product cycle, the better the traceability of the product. The use of blockchain technology allows the inclusion of all the information of the agricultural product throughout the supply chain, from its origin and cultivation properties to the details of shipping the product to the store; in order to verify the status of the food product \cite{BlockchainTechnologyAndTraceabilityAgrifood}.
    
    As mentioned by  Obeidat et al.  \cite{BlockchainTechnologyApplicabilityInTheTraceability}, the most significant advantage of using this technology is that all participants in the supply chain will be able to access the information directly, without manipulation of the data and with controllable access. With this information, important business decisions can be made, and operational efficiency can be improved, since it is possible to track whether the product was grown under the right conditions or if it encountered any problems during transportation. Today, large global companies such as Ikea have joined the blockchain technology \cite{ApplicationOfDigitalTechnologies} to ensure the end consumer that the product being purchased has accurate detailed characteristics on the label. 

    \subsubsection{Security} The implementation of a decentralized model for the storage of product information along the entire supply chain solves security and privacy issues thanks to data encryption without compromising important properties such as performance and scalability in the operations \cite{AdvancedSecurityModel}. Using IoT, data can be obtained by sensors and attached to the network automatically without the need for human intervention, so it will be accurate while respecting data integrity; additionally, when added to a blockchain block, it cannot be deleted or modified, so its storage is completely secure, reliable and immutable \cite{AComprehensiveReviewOfBlockchain}. 
  
    \subsubsection{Information Systems} Studies have shown that blockchain can be helpful for better implementations of information systems (IS). In China, blockchain technology has enabled governments to “track, monitor and audit” the supply chain while also allowing for more veritable manufacturing records \cite[p. 1360]{BlockchainAppFoodSupplyIS}. Further research, using grape production as a case study, proved that a blockchain-based IS not only reduced waste, but provided trust in the whole system through transparency and security throughout the supply chain \cite{BlockchainBasedDistributedCloud}.
  
    \subsubsection{Others} Although the category of “other” remains quite broad, we can further segment into three additional categories: 1) food safety 2) sustainable agriculture and the local economy and 3) agriculture finance \cite{sustainableSolutionsFood}. Blockchain technology can be a feasible solution to current food safety issues, although many barriers still exist that prevent widespread adoption of the technology \cite{appBlockchainTechFoodSafety}.
    
    Through the use of a ranking and award system, Casado-Vara et al. \cite{howBlockchainImprovesSC} promote a more sustainable agriculture practice by using smart contracts to remove intermediaries, leading to a more circular economy market. Blockchain has been used to create a Decentralized Employment System, where temporary workers can benefit from security in their short-term contracts \cite{BlockchainEmploymentContracts}. This attempts to solve the problems that can arise from a lack of trust during short-term employment tenures, common in the agricultural industry. 
    
    Agricultural financing that uses blockchain technology appears to be a relatively unexplored subject area. Given this, Pombo Romero and Rúas-Barrosa \cite{BlockchainFinancialInstrument} have presented a model that introduces decentralized financial instruments to provide funding to farmers pursuing photovoltaic irrigation systems (PVI). This aims to reduce the upfront cost required to implement sustainable agricultural practices such as the PVI. 

\subsection{Technical literacy within agriculture}
Throughout history, agriculture has been divided into four revolutions \cite{TheFourthAgricultural}. In the 1930s, research into the origins of agriculture began, and it was found that \textit{the First Agricultural Revolution} (AR) dates back more than 10,000 years ago in the Neolithic era when humans switched from foraging to farming \cite{AgricultureHistory}. There are several theories about the beginning of \textit{the Second AR}, for this Mingay \cite{SecondAgricultural} argues that it was not a single isolated event, but occurred during several moments between the 17th century with land repartition and the 19th century with the arrival of the Industrial Revolution in Great Britain. \textit{The Third AR}, also known as \textit{The Green Revolution}, began in the 1950s, mostly in Asia, producing more improved crops using chemical fertilizers and pesticides, reducing poverty in South Asia by 30 percent \cite{ThirdAgricultural}. In present day \textit{the Fourth AR} is beginning, including new technologies such as robotics, artificial intelligence (AI), IoT, global positioning systems (GPS) and blockchain technology \cite{PerceptionsOfTheFourthAgricultural}.

In the Fourth AR, farmers face several challenges that must be addressed. Although investment in new technologies brings greater benefits in the future, technological gaps prevent early adaptation. To achieve this, government intervention is needed to support farmers in making the agricultural transition. The Secretary of State for Environment, Food and Rural Affairs \cite{HealthAndHarmony} proposes the following main challenges for small and medium-sized agricultural enterprises: broadband coverage, mobile phone coverage, access to finance, affordable housing, availability of suitable business accommodation, access to skilled labour, transport connectivity.

One of the major agricultural producing countries is India due to its rapid economic growth in the last decades \cite{TheFourthIndia}. Lele and Goswami \cite{TheFourthIndia} present that Government of India is helping its people by using new reforms such as: 1) \textit{The Public Distribution System} in which the central government procures, stores, and transports food. 2) \textit{Liquefied Propane Gas} (LPG) subsidy by eliminating taxes on domestic LPG. 3) \textit{Mahatma Gandhi National Rural Employment Guarantee Scheme} that guarantees rural employment to its inhabitants for up to 100 days \cite{TheFourthIndia}. However, none of these include programs to help overcome the technological challenges mentioned above.

For all these reasons, the role of the state is essential to achieve growth in agricultural productivity. Technological growth in agricultural enterprises in developing countries depends on suppliers who are economically inaccessible; they do not have the economic capacity to invest in research and development and create their own technologies. Wanki \cite{TheLackOfDynamic} shows that the role of the state must be much greater in underdeveloped countries in order to combat poverty, but in developed countries it is the agro-input supply sector that must invest in R\&D in order to keep innovating their product line and remain competitive.

Lack of technical knowledge among participants is the single biggest barrier to successful blockchain implementation \cite{BarriersToAdoptionBlockchain}. In business supply chains, unfamiliarity is the second-biggest barrier to blockchain implementation \cite{BarriersAdoptionBlockchainBizSupply}. It is essential that this paper further investigate the cause of this much disruption in blockchain supply chain technology.

Numerous studies have been conducted to study the struggle of workers adapting to the fourth industrial revolution (industry 4.0). One example is explained by Huy et al. \cite{EducationComputerSkillsVietnamLaborers}, in which the Vietnam economy cannot take full advantage of the European Union-Vietnam Free Trade Agreement (EVFTA) signed in 2020. Their labourers, mostly agricultural workers, lack a level of computer skills necessary to contribute value following this trade agreement \cite{EducationComputerSkillsVietnamLaborers}. Yanzi et al. \cite{GlobalCitizensAwareness} further express this point by illustrating that not all countries can meet the education demands of Industry 4.0. In particular, developing countries with high levels of education inequality, such as Indonesia \cite{GlobalCitizensAwareness}.

Yanzi et al. \cite{GlobalCitizensAwareness} transition to explain the importance of digital literacy and the four impacts it has: 1) Increased hard skills 2) Enhanced global communication skills 3) increased soft skills 4) moral improvement. They broaden the definition of digital literacy to include the understanding of other cultures and world-views, with the goal to make global citizens \cite{GlobalCitizensAwareness}. 

To address this digital literacy deficit, society must develop lifelong learners: teaching how to continuously learn and build new skills \cite{LearningSkillsIndustry4.0}. It is important that this material be culturally conscious, diverse, and inclusive in nature \cite{LearningSkillsIndustry4.0}. Laar et al. \cite{Relation21CentSkillsDigitalSkills} reinforces this point by providing five contextual skills that must be developed in order to improve digital literacy: 1) ethical awareness 2) culture awareness 3) self direction 4) flexibility 5) lifelong learner. 

This paper will now analyse an industry in which there has been radical technical transformation: Healthcare. There will be a discussion around healthcare professionals current digital literacy and any solutions that have provided better technology integration among the workforce. 

At a university in Plovdiv, Bulgaria, a study was conducted to assess computer literacy among its healthcare students. The study found that knowledge was sufficient, although not outstanding \cite{compLiteracyHealthcareStudentsMedicalUniversity}. There was a call to host an informatics course, in which the students should build computing skills and build confidence and comfortability with computers \cite{compLiteracyHealthcareStudentsMedicalUniversity}. Kirkova-Bogdanova \cite[p. 807]{compLiteracyHealthcareStudentsMedicalUniversity} expressed the importance of this education by claiming that an increase in computer literacy is a “guarantee for a successful career”.

Another study conducted on 688 nurses found that computer literacy had a positive correlation on having a positive attitude towards computers \cite{nursesCompLiteracy}. This positive attitude had a positive correlation towards successful implementation of new computing technologies \cite{nursesCompLiteracy}. To put concisely, nurses must be better educated on computers to implement any effective computing technology \cite{nursesCompLiteracy}.

To better educate nurses, Vadillo found that computer training should be conducted in a classroom instead of 1 on 1 \cite{maximizingHealthProfessionUseComp}. Alongside classroom sessions, healthcare organizations should work towards creating a culture that values novel IT methods and encourages adoption of them \cite{acceptanceHealthITHealthProf}.

\section{Research Rationale}

One of today's major issues is food security, defined by the \textit{United Nations Committee on World Food Security} as a “mechanism in which all people, at all times, have physical, social, and economic access to sufficient, safe, and nutritious food that meets their food preferences and dietary needs for an active and healthy life” \cite{IFPRI}. Factors such as Global Warming, global conflict and the \textit{COVID-19} pandemic have led to an increase in world hunger \cite{WFP}. Focusing on climate change has resulted in decreased food production yields due to changes in temperature, precipitation and carbon dioxide \cite{AgronomicApproachToUnderstandingClimateChange}. Modern agricultural systems must adapt to these present-day problems, and here the Fourth AR plays a significant role. The use of these technologies contribute to improve the performance of agricultural production; however, it also has a great challenge, which is the access to all the infrastructure that this technology requires.

While digital transformation brings major benefits, adoption and accessibility seem to be the biggest issues. Since access to information requires very high upfront investment, Barrett and Rose \cite{PerceptionsOfTheFourthAgricultural} state that farmers who do not have the knowledge to adopt these new technologies are completely isolated from all the benefits they offer. For example, the paper Health and Harmony: the future for food, farming, and the environment in a Green Brexit \cite{HealthAndHarmony} state that there is a high barrier to adoption due to lack of knowledge and costs of the technology including financial issues and a lack of rural digital infrastructure.

Increased agricultural productivity brings as a benefit a reduction in food waste, thus contributing directly to food security. One of the most widely used technologies is precision agriculture, which uses different sensors to monitor the properties of crops, taking advantage of the data obtained to act in anticipation of factors such as diseases or insects and thus achieve a stable production \cite{TheRoleOfPrecisionAgriculture}. In this case, the role of blockchain can be very valuable, as data can be stored in a transparent, decentralized and scalable way, as this data can be stored and used from the beginning of the crop and throughout the entire supply chain \cite{IntegratingBlockchainAndTheInternet}. 

It is essential to investigate the level of technical literacy among farmers and identify whether this level is a barrier to widespread adoption of blockchain traceability technology. This paper will explore solutions that will bridge the gap between the complicated technical solutions and the users who will inevitably be using them.

Although several of the top supply chain organizations have previously chosen to adopt emerging technology, their hopes for the outcomes are higher than the outcomes themselves. According to Sodhi et al. \cite[p. 1]{WhyEmergingSupplyChain}, organizations frequently want short-term outcomes rather than medium and long-term results since “perceived benefits—and goals and constraints—depend on the interaction between technology and the users, not on the technology alone” \cite[p. 1]{WhyEmergingSupplyChain}. The ideas are highly general projects that do not satisfy the needs to tackle the unique difficulties of each organization, which is understandable given that the majority of innovations are outsourced. Spending time on the adoption of the technology and establishing concrete goals at every step of integration are two resources that might be helpful.

Farmers are not the only ones who face obstacles; businesses who operate with blockchain technology also encounter significant difficulties on a daily basis. Large volumes of data are handled in supply chain traceability that are vulnerable to assault, and even though the blockchain handles encrypted data, it can still be encrypted and leaked, violating client privacy  \cite{BlockchainTechnologyAndTraceabilityAgrifood}. Each piece of data included in the blockchain is also a transaction, which adds to the expenses that many users are unwilling to pay. For example, Khan and State \cite[p. 11]{LightningNetworkFees} make a comparison of transaction costs in 2019 between different blockchain such as Lightning Network, Raiden, Stellar and Bitcoin, their total cost is \$0.64, \$9.75, \$0.16 and \$0.32 respectively, suggesting that in large amounts of data transaction, the costs to be borne by the user will be very high. A possible solution to avoid these costs, would be to create cheaper consensus protocols; however, this may create gaps in the decentralization, security, scalability, and transparency of the blockchain.

As has been demonstrated, the use of technology is necessary to increase the yield of agricultural production, and it must also be implemented to have greater accessibility to the Fourth AR. This will contribute significantly to food security, increasing the yield of products, decreasing the cost of products and therefore decreasing hunger in the world. However, in cases where IoT systems are already used for agriculture, many have security and transparency issues that can be addressed using blockchain technology. Therefore, this proposed solution will suggest ways in which barriers to blockchain and IoT technology can be reduced and, with time, eliminated.

\section{Solution proposal}

This paper proposes the use of public policy and Research and Development (R\&D) to enhance the output of agricultural practices worldwide. The impact of \textit{public policy, public R\&D and private R\&D} in the agricultural sector will be discussed.

Moon \cite{LackDynamicCompetition} illustrates that public policy has the largest impact on the agricultural sector of developing countries, therefore the role of government in developing countries will be discussed. This paper proposes policy reform in the following areas: community development and legislation.

This paper suggests community development initiatives that focus on developing rural areas, which includes expenditure on modern technology and infrastructure, in addition to educational curriculum for agricultural workers. Modern skills and practices are essential for the highest level of agricultural productivity \cite{ChallengesAgricultureRuralDev} and this paper advocates that governments in developing countries need to make this a priority. In addition to supporting agriculture, expenditure in rural areas will grow industries and general economic value \cite{ChallengesAgricultureRuralDev}.

Ujo \cite{UJO} states that four conditions are required for agricultural and rural development:

\begin{itemize}
  \item Great equity in land distribution.
  \item Greater organization of producers and economic activities in rural areas.
  \item Better social relations and social services.
  \item Protection of rural interests by connecting them to the rest of the economy.
\end{itemize}

It is imperative that governments support rural economic activities through strong social support and connection to urban and higher populated areas.

Nchuchuwe and Adejuwo \cite{ChallengesAgricultureRuralDev} explain that the best way to support rural citizens is to “empower them through their occupations” \cite{ChallengesAgricultureRuralDev}[pg. 59]. Through continuous economic support, agricultural output, industry, and overall life quality can improve in rural areas.

This paper advocates for public investment into agricultural worker support. Economic progress will be achieved if investment is made into human capital, agricultural research, biophysical capital formation and rural institutions \cite{AfricanAgricultureStudies}.

Legislatively, governments of developing countries can pass laws to progress economic development. This paper proposes the protection of domestic agriculture workers through legislation reducing imports. The main challenge for developing countries is maintaining sustainable national growth in agriculture, given limited financial resources \cite{TradeLibStateSupport}. Low tariffs don’t protect local farmers in developing countries, it increases competition and decreases demand for their product \cite{TradeLibStateSupport}. 

Erokhin et al. \cite{TradeLibStateSupport} suggest subsidization of import substitution: replacing foreign goods with those locally produced. To further develop the local economy, governments can focus on geographic strengths to specialize and optimize production together with ensuring environment protection of agricultural production \cite{TradeLibStateSupport}. These are practices which prioritize the longevity of an agricultural ecosystem and understand the importance of supporting local farmers by increasing the demand for their product. 

This paper will now analyse the mechanism in which public R\&D is helpful for both developing and wealthy countries. The main way in which publicly financed R\&D supports development is through the contribution of knowledge to all society. In a case study from New Zealand, publicly researched foreign knowledge contributed greatly to their agricultural product growth \cite{AgricultureNewZealand}. If public institutions continue to publish research and develop new products, they are not only solving problems from their own countries, but global problems as well. Given this, public research is especially important for developing countries. Most agricultural R\&D and technology in developing countries is financed using public capital \cite{privitizationPublicPolicy}.

In addition to advocating for public R\&D as a whole, this paper advocates for public investment in basic and basic-applied research. This type of research increases the level of private R\&D in a given country \cite{privitizationPublicPolicy}. Conversely, public investment in developmental research decreases the level of private R\&D \cite{privitizationPublicPolicy}. It is essential that governments set a foundation by investing in basic and basic-applied research and encourage private industries to finance the development of products. 

Finally, there is private R\&D, which is necessary to maintain a high standard of quality in the agricultural industry. Generally, these private investments are seen more in developed countries, where public R\&D provided public policies in the past to boost the industry \cite{TheLackOfDynamic}. It is important that the state can support the private sector by providing funding so that it can continue to conduct research and keep up to date to meet the new challenges of Fourth AR. 

The entire food industry invests in R\&D to improve its production processes and product yields. This research evaluates cultivation procedures, biochemical inputs such as pesticides and chemical fertilizers, machinery, and technology in order to boost global sales \cite{TheGrowingRolePrivate}. For example, Fuglie et al. \cite{TheGrowingRolePrivate} conducted research on the role of the private sector for the agricultural industry and argues that “Global private spending on agricultural R\&D (excluding R\&D by food industries) rose from \$5.1 billion in 1990 to \$15.6 billion by 2014”. With this it can be concluded that the private sector, is making the relevant investments in order to be updated day by day.

The private sector in R\&D presents additional challenges to economic investment. Pray and Fuglie \cite{AgriculturalResearchByThePrivate} conducted research in which it found that intellectual property rights have created large gaps to invest in R\&D, considering that patents do not allow access to research for its development. To solve this problem, technological public policies must be created to support industries in the correct use of new technological advances and thus allow them to participate.

\section{Strengths and Weaknesses}

There are several strengths that will result from government and corporate investment in emerging blockchain technologies and their applications. Firstly, less food waste along the supply chain. A case study by Lucena et al. \cite{CaseStudyGrainQuality} illustrates a 15\% increased valuation in Brazilian grain exports when using blockchain technology. This increase in efficiency results in greater retention of food and will increase food security globally.

In addition, any increase in food supply will also have economic benefits. The law of supply and demand claims that as the supply of a good increases, the price of that good will decrease \cite{LawSupplyDemand}. We can therefore conclude that blockchain technology reducing food waste will act as an agent fighting against inflation and the rising cost of food. 

Along with less food being wasted, blockchain technology increases food safety \cite{blockchainEconomicsFoodSafety}. Increasing information accessibility along the supply chain enables stronger food safety protocols from all parties involved and health benefits for the customers of that food \cite{blockchainEconomicsFoodSafety}. Additionally, blockchain technology enables stronger crisis management. If contaminated food products are detected, it is easier to find the sole problem and eliminate this contamination effectively instead of removing the entire product line \cite{blockchainEconomicsFoodSafety}.

Improvements in traceability technology will also enable for a better consumer experience. The consumer will gain a better understanding of what they are purchasing and where it came from. Businesses, through the use of blockchain technology, will benefit through increased sales growth and retention of customers \cite{blockchainEconomicsFoodSafety}.

There can be additional strengths if governments of developing countries change their policy to better protect local farmers. By decreasing international trade, these governments can create local agricultural development that is sustainable over time \cite{TradeLibStateSupport}. Through the increased production of local food, diets will maintain a variety of nutrients and if a food disaster were to occur: the local economy would be better equipped to manage it \cite{blockchainEconomicsFoodSafety}.

There appear to be three main weaknesses, or challenges, in achieving accessibility to public and private promotion of blockchain technology. Implementation will be challenging as the agricultural supply chain is extremely complex. There are greatly diverse stakeholders stretching across many countries. Laws, languages and processes will vary, bringing opposition to uniformity in a solution. 

As mentioned in the literature review, a lack of technical literacy is a large barrier to the use of blockchain technology. In business supply chains, unfamiliarity with technology is the second-biggest barrier to implementation of blockchain technology \cite{BarriersAdoptionBlockchainBizSupply} Widespread education on these processes will be required before widespread adoption may occur.

Further opposition can be found in the cost of implementation. Although upfront costs can be very high \cite{AnalysisBlockchainProsCons}, over time a blockchain solution will be cheaper to manage than a centralized database. Over the course of time, the cost of these technologies will likely decrease, allowing for adoption in smaller companies \cite{blockchainEconomicsFoodSafety}.

\section{Discussion and future research agenda}

It is clear that technology adoption is a major challenge for farmers and this affects the end of the supply chain, i.e. the final consumer, the most. However, the problem goes beyond adoption as there are several important factors such as research, development, and project financing. All the gaps need to be filled in order to level the playing field and make progress in the Fourth AR. It is proposed to continue in-depth research in the following areas:

\subsection{What are the R\&D on emerging technologies adapted to blockchain in the supply chain?}
As mentioned in the Literature Review, there are many other fields besides traceability in which blockchain brings value. It is important to examine the new applications of this revolutionary technology, verify its scalability and point out which are the most important and feasible to adapt in the short term, thus, favouring the producer and the consumer. 

\subsection{How can farmers reduce the lack of technological and financial know-how?} 
Many agribusinesses have not been able to join the Fourth AR because of the large gap that exists; there are strong preferences to continue doing what they know, and they do not know the benefits of technology. Extensive research should be done on the creation and development of programs (state and consulting) to help agribusinesses adopt new technologies.

\subsection{Is there an impact of new technology regulations on blockchain?}
Global governments are creating new regulations on technology, and lately on blockchain because of its anonymity issues, which may affect both R\&D and early adoption by new users. It is necessary to review a comparison between countries with more technological development and compare which have more incentives to generate new developments.

\subsection{How is the Fourth AR being leveraged to address food insecurity issues in the 21st century?}
More developed countries can help push developing countries to adopt new technologies more quickly. Research on knowledge transfer between countries should be conducted and incentives for further R\&D in the public and private sector should be analysed.

\section{Conclusion}

It is necessary to apply knowledge to the great problems of the twenty-first century, such as food safety and food security, so it is critical to adapt quickly to the new technologies brought by the Fourth AR to achieve better production performance, which benefits both the producer and the consumer. The agricultural industry is using IoT to revolutionize its supply chain system; however, it has some transparency and centralization flaws that can be addressed by using blockchain technology. An analysis of the main applications in which blockchain is important to develop and execute was conducted to identify the impediments to implementing these technologies.

Various ARs throughout history demonstrate the importance of implementing new production methods to improve product quality. Adapting to the Fourth AR, on the other hand, is difficult given the difficulty of adopting the technology due to misinformation and costs. Developing countries are the most affected because there is insufficient government support to encourage the promotion of these new revolutionary technologies; however, in developed countries, the public sector must face the challenges of R\&D to stay current. Furthermore, there are barriers for technology companies, such as the handling of large amounts of private data and their vulnerability to attacks, as well as high development costs and infrastructure upgrades.

This article discusses the impact of public policy, public R\&D, and private R\&D on these technology implementation issues. Depending on the country's income, the public, or private sector should have a greater impact, so that both farmers and technology consulting firms can drive the Fourth AR. The benefits of implementing blockchain technology along the supply chain include reduced food waste, increased sustainable production, lower costs, increased product information, and greater support for local production in each country; however, some drawbacks were discovered, including implementation complexity, a lack of technical literacy, and the costs of these technologies.

The adoption of new technology is a large worldwide problem, but it will have a significant impact in the medium and long terms. By integrating blockchain technology into various supply chain applications, it will be easier to produce more food while reducing security and transparency gaps. The challenges must be solved from the public and private sector for greater equality in the distribution of knowledge that will strengthen the Fourth AR.

\bibliographystyle{IEEEtranDOI}
\bibliography{referencesDOI}

\end{document}